# Soliton Stability and the Foundations of Physics: A Challenge to Real Analysis and Numerical Calculation

Paul J. Werbos[1]
ECCS Division, National Science Foundation

## Abstract

At present, there exists no physically plausible example of a quantum field theory for which the existence of solutions has been proven mathematically. The Clay Mathematics Institute has offered a prize for proving existence for a class of Yang-Mills theories defined by Jaffe and Witten. This paper proposes a multi-stage strategy for proving existence for a quantum version of the "'tHooft Polyakov" (tP) field theory, and argues that this theory – while not renormalizable or physically plausible as it stands – opens up a clear path to a physically plausible well-defined theory. The key initial challenge is to first prove stability for a classical version of this theory, in the spirit of Walter Strauss. The stability results of Bogomolnyi for classical PDE systems are widely cited as a foundation of string theory, but they leave key questions unaddressed, and may even call for small modifications of the tP model itself. This paper calls attention to the questions which require more attention, and proposes some partial possible answers. It also specifies a new numerical method to check for stability for the usual "hedgehog" types of solution, a method which may have broader potential uses for any ODE problems where accuracy in derivatives is as important as accuracy in function values.

## 1. Introduction and Plan

This paper will mainly focus on the following question: do stable "soliton" solutions actually exist for the PDE which emerge as the Lagrange-Euler equations for the following Lagrangian:

$$\mathcal{L} = \tfrac{1}{4} G^{\mu\nu a} G^a_{\mu\nu} - \tfrac{1}{2}(D_\mu Q^a)(D^\mu Q^a) - \tfrac{1}{8}\lambda(Q^2 - F^2)^2, \qquad (1)$$

where the underlying fields are $Q^a$ and $A^a_\mu$ for a = 1,2,3 and μ=0,1,2,3, where λ and F are parameters, and where we have used the definitions:

$$G^a_{\mu\nu} = \partial_\mu A^a_\nu - \partial_\nu A^a_\mu + e\varepsilon_{abc} A^b_\mu A^c_\nu \qquad (2)$$

$$D_\mu Q^a = \partial_\mu Q^a + e\varepsilon_{abc} A^b_\mu Q^c \qquad ? \qquad (3)$$

I will refer to this as the 'tHooft-Polyakov model, or the "tP" model. So far as I know, this specific model was first proposed by 'tHooft[1], who noted that it belongs to a type of model discussed by Georgi and Glashow in 1972. 'tHooft[1] and Polyakov[2] made this model famous in 1974, when they showed that this system possesses equilibrium states which appear to possess magnetic charge. Here I will mainly use the notation of Hasenfratz and 'tHooft[3] in describing the system.

---

[1] The views expressed here are those of the author, not those of his employer; however, as work produced on government time, it is in the "government public domain." This allows unlimited reproduction, subject to a legal requirement to keep the document together, including this footnote and authorship. Some related material is posted at www.werbos.com.

Unfortunately, I cannot provide a decisive answer to this question at the present time. I hope that true experts in real analysis will be able to follow up.

Section 2 will discuss the importance of the question to mathematics and dynamical systems. Section 3 will discuss the importance to physics, and the possible path to the larger goal of proving existence for a physically plausible model. Section 4 will discuss what I can say about the question itself, and various subquestions which cry out for further investigation. Finally, the appendix will specify a new family of numerical methods which may be used in these studies, or on a broader class of 1-D problems.

## 2. Mathematical Importance of the Question

Do stable "solitons" exist for any system of partial differential equations defined by a Lagrangian $\mathcal{L}(\varphi(x_\mu), \partial_t\varphi(x_\mu), \nabla\varphi(x_\mu))$, where $\varphi$ is a set of n continuous functions or fields over $R^n$, functions of points $x_\mu$ in Minkowski space? This is a very fundamental question, for which the answer is not yet known. Note that $\varphi$ here is a vector in the mathematical sense, not necessarily a covariant vector in the sense of relativity; in effect, it is shorthand for "a bunch of stuff," which could be a combination of scalars, vectors, tensors, and spinors, etc., *but not* unitary matrices $U(\mathbf{x})$, because the components of $U(\mathbf{x})$ are not free to vary over all of $R^n$.

If we can prove that the tP model, or a slight modification of it, possesses stable "solitons," then we can answer the question above as "yes." It would be a case of proof by example.

The most complete discussion of this question to date is in the book by Makhankov, Rybakov and Sanyuk (MRS)[4]. MRS provide stability arguments for the Skyrme model, which relies on unitary matrices $U(\mathbf{x})$. Theories which allow the field components to range over all of $R^n$ are what they call "topologically trivial." They propose a "Generalized Hobart-Derrick Theorem," which asserts that stable solitons cannot possibly exist in such theories. If Bogomolnyi's arguments could be translated into a rigorous proof of stability, this would disprove the claim of MRS, or at least force a restatement of the claim; however, this has not yet been done.

Before trying to prove that stable solitons exist in this dynamical system, we need to define what we mean by a "stable soliton." Unfortunately, the word "soliton" has been used to refer to different things in physics and mathematics. In this paper, I will use the word "soliton" to refer an equilibrium state which meets the requirements to be chaotic soliton or "chaoiton" as defined in a previous paper[5]. This basically means that the equilibrium state is stable with respect to small perturbations, except that they may move the state a small distance or even impart some velocity or angular momentum to it. For a state of positive energy, in a theory with positive definite energy near the vacuum state, it is enough that the state be a local minimum of energy , over all possible small perturbations.

The well-known Hobart-Derrick Theorem (discussed by MRS[5]) implies that solitons are impossible in any quasilinear PDE system over Minkowski space. It is proven very simply[5] by considering perturbations in which $\underline{\varphi}(\mathbf{x})$ is replaced by $\underline{\varphi}(\lambda\mathbf{x})$, where λ is close to one. It is encouraging that this argument does not go through directly for the tP model.

## 3. Importance of the Question for Physics

The problem statement by Jaffe and Witten[6] for the Clay Mathematics Institute states: "…one does not yet have a mathematically complete example of a quantum gauge theory in four-dimensional theory in four-dimensional space-time, for even a precise definition… Existence theorems that put QFTs on a solid mathematical footing are needed ..it is a great challenge for mathematicians to understand the physical principles that have been so important and productive throughout the twentieth century." The most recent evaluation of the status of this challenge from the Clay Institute[7] states: "Perhaps the most basic mathematical question here is to specify a class of initial conditions for which we can guarantee existence and uniqueness of solutions." These challenges in turn are basically just a steppingstone to the larger challenge of finding a *physically plausible* quantum field theory for which one can guarantee existence of solutions.

Since no one has met any of these challenges as yet, there is no proven strategy for meeting them. Here I would like to suggest the following *plausible* multi-stage strategy:

(1) First prove soliton stability (soliton existence) for the tP model, or a slightly modified version of it – as discussed in section 4 and the Appendix.

(2) Use the insights from this stability proof to deduce bounds on the fields in this system – in effect, to understand why it does not "blow up" into discontinuous states;

(3) Based on these bounds, prove the existence of solutions for the classical PDE, using the same general approach which has worked for the Yang-Mills system by itself[8];

(4) Use the generalized "P" mapping of coherence theory[9-11] to map the energy density of the tP model into the equivalent normal form Hamiltonian H. This is possible, because $Q^a$ and $A_\mu^a$ are both bosonic fields. Within Fock-Hilbert space, use the P-mapped version of the classical distance metric instead of the usual $L^2$ metric. (This avoids the difficulties which come when a soliton plus a photon of epsilon energy are treated as something very far away from the soliton.) Exploit the results from (3) and the P equivalence relations to prove existence of solutions for the "many worlds" dynamical system:

$$\dot{\psi} = iH\Psi \qquad (4)$$

for this choice of H.

(5) Having met a key part of the Clay challenge, then move on to a more realistic model, perhaps by replacing $A_\mu^a$ by the vector bosons of electroweak theories, as previously proposed[12], or perhaps by introducing a second topology term to fix electric charge, or both.

Of course, with a multi-stage strategy, we would always want to be ready to make midcourse corrections as we learn more at each stage.

A few additional comments are in order for stages 4 and 5, even though they are not the main topic of this paper.

With regard to stage 4 – there are many possible forms of quantum field theory, none of which have met the Clay challenge as yet. I am proposing that we try for a many-worlds theory, in part because it allows the use of strategy proposed here, and in part because many worlds theories do better so far in confronting empirical reality in the world of engineering based on quantum electrodynamics[13]. Appendix B gives some thoughts about how to make the connection from the classical fields to the quantum fields – but the initial challenge here is how to better establish the properties of the classical field theories themselves.

With regard to stage 5, the goal is basically to derive a "bosonic standard model[14,15]." It is important to remember that the ultimate field equations need not include separate components for each possible type of particle. With a relatively few truly stable solitons, it is possible to have bound states (like hadrons) and even bound states of bound states, and very complex and esoteric emergent behavior. But before we can settle on the right theory, we need a wider class of theories to choose from, and before that we need to have at least one example.

It is also important to remember that there is no requirement in mathematics that a valid dynamical system must be *renormalizable*. In fact, it is even possible to be "finite" without being renormalizable! It is well-known[16] that the masses predicted by the Feynman path quantization of the tP model are in fact finite, to all orders of quantum correction; infinite mass corrections through renormalization are not needed. Methods like WKB can work even in cases where the usual perturbation about the vacuum state do not[16]; it is an act of incredible hubris to assume that nature can only possibly be governed by dynamics which fit with the first approximation scheme we happen to try on it. Furthermore, the usual sorts of variational (energy minimization) methods used in atomic and molecular physics, to calculate energy levels, could also be applied to the prediction of the masses of the hadrons, an important empirical issue where well-defined quantum field theory has not yet been able to do as well as phenomenological models[12], as in the work of Paolo Palazzi and of MacGregor.

After one has a bosonic standard model, it should be possible again to exploit classical-quantum equivalence, and directly use the metrification procedure given by Carmelli[17] in order to unify that model with general relativity.

### 4. Reassessing Stability of Solitons in the tP Dynamical System

### 4.1 Summary of Bogomol'nyi's Analysis

Bogomol'nyi's classic paper[18] is the primary source for what is known about the stability of magnetic monopole solutions in the tP system. (Relevant work by Striet and Bais[19,20] and Affouf[21] will be discussed in section 4.2.)

Bogomol'nyi discusses the issue of stability for equilibrium states for which $A_0^a=0$ in the tP system. He defines "stability" as being a state of minimum energy (not necessarily a unique minimum) *considering only* the energies of other states of the same total magnetic charge.

For any equilibrium state with $A_0^a=0$ everywhere, he states that the total energy equals:

$$E = \int d^3x \left\{ \frac{1}{4}(G^a_{mn} - \varepsilon_{mnp}D_pQ^a)^2 + \left[ \frac{\varepsilon^{mnp}G^a_{mn}D_pQ^a}{2} \right] + \frac{\lambda}{8}V(Q) \right\} \quad (5),$$

where I have introduced the function "V(Q)" to represent the usual Higgs term in the 'tHooft-Polyakov model. Equation 5 corresponds to Bogomol'nyi's equation 5.3. His version of the Higgs term varies somewhat from equation to equation in his paper, but does not affect his arguments. Going back to 'tHooft[3], I will assume:

$$V(Q) = (Q^2 - F^2)^2 \quad (6)$$

He then deduces that the integral of the term in square brackets is actually constant for all field states which have the same magnetic charge; thus it is enough to show that we can find a global minimum for

$$E' = \int d^3x \left\{ \frac{1}{4}(G^a_{mn} - \varepsilon_{mnp}D_pQ^a)^2 + \frac{\lambda}{8}V(Q) \right\} \quad (7)$$

over states of the desired magnetic charge. He then argues that we can find solutions of the desired magnetic charge which obey:

$$G^a_{mn} = \varepsilon_{mnp}D_pQ^a \quad (8)$$

Equation 8 is so important that it deserves a name; I will call it the "adiabatic potential equation." Bogolmonyi then shows how we may find solutions of the adiabatic potential equation by inserting a normalized version of 'tHooft's ansatz[3]:

$$Q^a = r_aQ(r) \quad (9)$$
$$A^a_i = \varepsilon_{ial}r_lW(r) \quad (10)$$
$$A^a_0 = 0 \quad (11)$$

which leads to a set of coupled ordinary differential equations. In the limit as λ goes to zero, he argues that this sets equation 7 to zero. Since equation 7 is nonnegative, a value of E' of zero must be a global minimum, and therefore a stable state of the system in three dimensions.

**4.2 Reassessment of the Analysis**

If the argument in section 4.1 could be elevated to being a theorem, it would be enough by itself to fulfill the goal of this paper (goal number 1 in the list in section 3). But is it possible?

Most of the steps should be easy enough to formalize. For solutions where Q has interesting topology (magnetic charge) in the limit as r→∞, it is necessary to prove that small perturbations cannot change that magnetic charge. This seems straightforward enough.

As a pedestrian exercise, I have gone ahead and inserted equations 9 through 11 myself into equation 9, and verified that it does yield two coupled equations similar to (and presumably equivalent to) Bogmol'nyi's, if I exploit the identity:

$$r_m \sum_l \varepsilon_{nal} r_l - r_n \sum_l \varepsilon_{mal} r_l = \varepsilon_{mna} - r_a \sum_l \varepsilon_{mnl} r_l \tag{12}$$

Existence of smooth solutions to the two coupled equations (first order ODE) should be easy enough to prove. (At any point in space **x**, the three-dimension vector of $r_a$ (as a=1,2,3) is just the vector **x**/r = **x**/|**x**|.)

But then comes the problem. Because we can set the first term in E' to zero (its global minimum), Bogomolnyi asserts that this constitutes a global minimum in the limit as $\lambda$ goes to zero. That is as far as the argument goes. Others have cited this argument "in the limit" as a reason to believe that we always have stability in this system. Furthermore, it is well known in mathematics that a system can have bounded, well-defined behavior for any value of a parameter $\lambda>0$, but radically different behavior when $\lambda$ actually equals zero; things may often go to infinity as $\lambda$ goes to zero. Here, where the very existence of the soliton depends on the strength of the Higgs term, we should actually *expect* a qualitative change in behavior in the case where $\lambda=0$. Thus there is strong reason to reconsider the analysis.

For the case where $\lambda>0$, let us look again a second time at the adiabatic potential equation, equation 8. For the case of solutions which obey the 'tHooft ansatz, we do not have much freedom in solving this equation. But in the general case, in three dimensions, even if $Q^a$ obeys equation 9, would it be possible to find values of the fields $A_\mu^a$ which solve equation 8 for any choice of Q(r) which meets the boundary conditions (and a few reasonable smoothness conditions)? Starting from field values at all points at some radius r, equation 8 would appear to be a well-posed system for finding field values for $A_\mu^a$ at larger radii r which can accommodate Q(r), even though these field values would not fit equation 10.

More precisely, if we substitute the definitions from equations 2 and 3 into equation 8, and (like Bogomolnyi) consider only solutions where $A_i^0=0$, then the adiabatic potential equation becomes:

$$\partial_m A_n^a - \partial_n A_m^a + e\varepsilon_{abc} A_m^b A_n^c = \varepsilon_{mnp}\left(\partial_p Q^a + e\varepsilon_{abc} A_p^b Q^c\right) \tag{13}$$

Formally, this is a system of 27 equations (for all choices of m, n and a from 1,2,3); however, for any choice of a, there are really only three equations present, because when m=n the equation is always true, and because an interchange of m and n yields equivalent equations. In summary, we have a system of only nine equations; it should be possible to develop a path for the nine variables $A_n^a$ to satisfy these nine equations, for any smooth enough "source" function $Q^a(r)$. Let us define the *adiabatic potential hypothesis* as the hypothesis that this is indeed the case.

*If the adiabatic potential hypothesis is true*, then for $\lambda>0$ the Bogomol'nyi solution is not a solution of minimum energy (since $V_Q>0$). It that case, there exists a *family* of states which goes to E'=0 in a limit of the family, but the limit of the family is not a smooth solution. To define that family, we attach $Q^a$ to its asymptotic limit, where V(Q)=0, for all points r outside some small radius $r_0>0$, while still keeping $Q^a(0)=0$ for

r<$r_0$. As we make $r_0$ closer and closer to 0, we bring E' closer and closer to its true global minimum. The problem is that the limit of this series is not a smooth function, but a singular function. This proves that equation 5 and equation 7 do not in fact have a global minimum within the (open) set of smooth, differentiable functions, *if $\lambda>0$ and if equation 8 has these more general solutions*.

In summary, it is not at all clear as yet whether a stable solution really does exist for the tP system for the case where $\lambda>0$.

A recent careful study[19,20] of "Alice Electrodynamics" (a close relative of 'tHooft-Polyakov and of string theory) has shown numerically that the usual radially symmetric "soliton" solutions are in fact unstable. They argue, on intuitive grounds, that lower energy solutions with a more complex topology may be stable; however, based on the argument above, that is not at all obvious, at least for the 'tHooft-Polyakov model and its relatives. Their main finding is that the usual solutions are unstable with respect to perturbations which *are not within the usual ansatz (equations 9-11)*. Good numerical solutions are indeed an important tool in proving *instability* of a proposed soliton state.

Fortunately, there probably is an easy fix here, if this problem turns out to be real. It may not be esthetically appealing to those who insist that the universe must be completely beautiful, but it does appear to work. We can simply modify the Lagrangian of the 'tHooft-Polyakov model by adding a small term -$\eta(\partial_\mu Q^a)^2$, which changes the energy to

$$E = \int d^3 x \left\{ \frac{1}{4}(G_{mn}^a - \varepsilon_{mnp} D_p Q^a)^2 + \left[\frac{\varepsilon^{mnp} G_{mn}^a D_p Q^a}{2}\right] + \frac{\lambda}{8}V'(Q), \right\} \qquad (14a)$$

where:

$$V'(Q) = V(Q) + \eta(\nabla Q)^2 \qquad (14b)$$

If $\eta$ is extremely small, this could have vanishingly small large-scale observable consequences. But when we minimize V', instead of minimizing V, the $\eta$ term makes this minimization task equivalent to the familiar task of solving the simple Schrodinger equation of elementary quantum mechanics, for a spherically symmetric potential. Bogmol'nyi's arguments would still go through, except that my extension of those arguments results in a smooth minimizer of energy. In the limit as $\eta \to 0$, this approaches the singular soliton of the 'tHooft-Polyakov model proper, but for any $\eta>0$ it is still finite. Of course, this procedure could be applied to other models in the same family, such as those discussed in section 3.

In the end, taking the limit as $\eta \to 0$ may lead us to something very similar to the usual standard model in its behavior. Yet it would have the exciting property of concentrating charge in an extremely small region of space, without requiring a huge negative energy in that region to cancel out the positive energy in the electromagnetic field outside of that region. This could be seen as just another way to *do* regularization. But in this case, each value of $\eta>0$ is in itself a finite theory, for which we can probably prove rigorous existence theorems. It is a major step towards the larger goal of getting back to reality.

### 4.3. Some additional comments on stability

To prove stability against *all* small perturbations, we need to consider perturbations which may not be spherically symmetric. But the change in energy, $\delta^2 E$ or $\delta^2 H$ (discussed at length in previous papers[4,22]), obeys something very similar to the Schrodinger equation of elementary quantum mechanics, in which the static field values determine the "potential well" for the perturbation. When the energy operator is simple enough, there is a separation of coordinates, exactly like what we see in atomic physics; thus the lowest energy perturbations are in fact radially symmetric, and it is enough to consider stability against perturbations which are radially symmetric.

The problem here is that we *do know of* a nontrivial perturbation which will always have zero energy. Consider a small *translation* of the soliton in 3-D space! This corresponds to perturbation of the form f sin θ, which is not radially symmetric; it is part of the first spherical harmonic space of functions. Because of this mode, we can automatically deduce that any field theory which obeys such a simple separation of coordinates must posses perturbations of negative energy, and therefore not allow stable solitons. (Or at least not spherically symmetric ones.)

Intuitively, the obvious way to overcome this obstacle is to couple spin with isospin, such that the separation of coordinates results in letting us continue to work in the $Q^a(r)$ ansatz. That's essentially what the gauge invariant part of the tP model buys us, and why this works.

The comments in this subsection do not override the analysis of section 4.2. Rather, they are intended to provide some intuitive explanation of why it works, and to suggest the possibility of more general mathematical results in future research.

Rajaraman[16] has cited important papers which provide additional ideas for how to consider solutions and issues beyond the usual spherically symmetric case.

If the adiabatic potential hypothesis is true, then of course the η term in equation 14 plays an essential role in establishing the bounds which make it possible to prove the existence of solutions. If not, then we still do not know whether the known solutions to the original tP model (with λ>0) are stable or not. Numerical work can also be of value in helping us to find out.

# **References**

1. G. tHooft. Magnetic monopoles in unified gauge theories. *Nucl. Phys.*, B79:276-284, 1974.

2. A.M. Polyakov. Particle spectrum in quantum field theory. *JETP Lett.,* 20:194-194, 1974.

3. P. Hasenfratz & G. 't Hooft, Fermion-boson puzzle in a gauge theory, *Phys Rev Lett*, Vol. 36, p. 1119-1122 (10 May, 1976)

4. V.G.Makhankov, Yu.P.Rybakov and V.I.Sanyuk,*The Skyrme Model*, Springer-Verlag, 1994.

5. P.Werbos, Chaotic solitons in conservative systems: can they exist?, *Chaos, Solitons and Fractals*, Vol. 3, No.3, p.321-326,1993.

6. Jaffe and E. Witten , "Quantum Yang-Mills Theory," in The Millennium Prize Problems, J. Carlson, A. Jaffe, and A. Wiles, Editors, pages 129—152, American Mathematical Society, Providence, 2006
http://www.claymath.org/millennium/Yang-Mills_Theory/yangmills.pdf

7. Michael R. Douglas, Report on the Status of the Yang-Mills Millennium Prize Problem, April 2004,
http://www.claymath.org/library/annual_report/douglas_quantum_yang_mills.pdf

8. Walter A. Strauss, *Nonlinear Wave Equations*, CBMS Regional Conference Series in Mathematics Number 73, American Mathematical Society, 1989

9. K.E. Cahill and R.J. Glauber, Ordered expansions in boson amplitude operators, *Phys. Rev.* $_{Vol.\ 177}$p. 1857 (1969)

10. L. Mandel and E. Wolf, *Optical Coherence and Quantum Optics*. Cambridge U. Press (1995), chapter 11.

11. P. Werbos, Classical ODE and PDE Which Obey Quantum Dynamics, *Int'l J. Bifurcation and Chaos*, Vol. 12, No. 10 (October 2002), p. 2031-2049. Slightly updated as quant-ph 0309031.

12. P. Werbos, *Schwinger's magnetic model of matter – can it help us with grand unification*, revised January 2008, arXiv:0707.2520

13. P. Werbos, Bell's Theorem, Many Worlds and Backwards-Time Physics: Not Just a Matter of Interpretation, *Int'l J Theoretical Physics*, April 2, 2008 (epub date), arXiv: 0801.1234.

14. Tanmay Vachaspati, On a dual standard model. Hep-ph 0204106.

15 Levon Pogosian, Daniele A. Steer, Tanmay Vachaspati, Triplication of SU(5) monopoles**.** *Phys.Rev.Lett*. 90 (2003) 061801

16 R. Rajaraman, *Solitons and Instantons: An Introduction to Solitons and Instantons in Quantum Field Theory*. Elsevier, 2$^{nd}$ edition, 1989.

17. M. Carmelli, *Classical Fields: General Relativity and Gauge Theory*. Wiley (1982), section 3.3

18 E.B. Bogomolnyi, The stability of classical solutions. *Sov. J. Nucl. Phys.*, Vol. 24, No. 4, October 1976, p. 449-454

19. F.A. Bais and J. Striet. On a core instability of 'tHooft Polyakov type monopoles. *Phys. Lett.*, B540:319-323, 2002.

20. J. Striet and F.A.Bais, More on core instabilities of magnetic monopoles. arXiv: hep-th/0304189v2, June 2003.

21. M. Affouf, Numerical scheme of magnetic monopoles. *Proc. WSEAS Int'l Conf. on Applied Mathematics*, Dallas, Texas, 2007.

22 . P.Werbos, New Approaches to Soliton Quantization and Existence for Particle Physics, http://arXiv.org/abs/patt-sol/9804003

23. B. Julia and A. Zee, Phys. Rev. D **11**, 2227 (1975)

24. Tai Tsun Wu and Chen Ning Yang, Some solutions of the classical isotopic gauge field equations. In *Properties of Matter Under Unusual Conditions*, H. Mark and S Fernbach eds., Wiley Interscience 1968.

25. M.K. Prasad and C.H. Sommerfeld, *Phys. Rev. Lett.*, 35, p.760, 1975.

26. M.H. MacGregor, *A "Muon Mass Tree" with α-quantized Lepton, Quak and Hadron Masses*. arXiv: hep-ph/0607233

27. Malcolm H. MacGregor, *Power of α: Electron Elementary Particle Generation With α-quantized Lifetimes and…*, World Scientific, 2007

28. Paolo Palazzi, *Particles and Shell.* CERN-OPEN-2003-006. arXiv: physics/0301074.

## **Appendix A. Numerical Approaches to Testing Stability**

One way to follow up on the analysis of this paper is to do some numerical experiments. Of course, it is much easier to do numerical experiments on fields which vary as a function of one spatial coordinate, r, than functions which vary as a function of r and z (as in Striet and Bais[19,20]) or as a function of $x_1$, $x_2$, and $x_3$. The initial work by 'tHooft, Polyakov and Bogomol'nyi all focused on particular static solutions. More precisely, 'tHooft[3] proposed an "ansatz," a particular family of possible solutions for $Q^a$ and $A^a_\mu$ based on functions Q(r) and W(r), shown in equations 9-11. Julia and Zee[23] performed numerical experiments based on a more generalized ansatz using three functions H(r), J(r) and K(r), based on earlier work[24] by Wu and Yang (whose collaborations with Mills and Lee are very well-known). In the notation used by Prasad and Ju, this ansatz is:

$$A_i^a = \frac{\varepsilon_{aij} x^j (1 - K(r))}{er^2} \tag{15}$$

$$A_0^a = \frac{J(r)}{er^2} x^a \tag{16}$$

$$(Q^a =) \phi^a = \frac{H(r)}{er^2} x^a \tag{17}$$

Note that "$x^a$" refers to the a-th coordinate of the vector **x**, and that equation 17 is telling us the proposed value of $Q^a(\mathbf{x})$, which is 'tHooft's notation for what Julia and Zee would call $\phi^a(\mathbf{x})$.

Thus one way to follow up on the analysis above is to use the H/J/K ansatz to construct solutions to the tP model, and then test them to see whether they truly are local minima with respect to small perturbations of the functions H, J and K. If they are not, this by itself would prove that they are not stable. If they are truly local minima with respect to these perturbations, it would not prove stability, but it would at least open the door to that possibility. It would be a first step towards a test of stability against more general perturbations.

The first step in doing *that* work is to construct solutions to H(r), J(r) and K(r) numerically. Following the elegant presentation of Prasad and Sommerfeld[25], this means finding values of these functions which "minimize":

$$E = \frac{4\pi}{e^2} \int_0^\infty dr \left( K'^2 + \frac{(K^2-1)^2}{2r^2} - \frac{J^2 K^2}{r^2} - \frac{(rJ'-J)^2}{2r^2} + \frac{H^2 K^2}{r^2} + \frac{(rH'-H)^2}{2r^2} - \frac{\mu^2 H^2}{2} + \frac{\lambda H^4}{4e^2 r^2} \right) \tag{18}$$

For my proposed modified model, we would add the following term to the integrand:

$$+\eta' \left( \frac{H'}{r} - \frac{1}{r^2} \right)^2 \tag{19}$$

Following Julia and Zee, we may substitute $\beta^2 = e^2/\lambda$ and assume $\mu=1$; thus the solutions here depend only on the choice of $\beta$ (and on $\eta'$) and on the boundary conditions. Like Julia and Zee, we may use the boundary conditions:

$$J(0) = H(0) = K(0) - 1 = 0 \tag{20}$$

But we still face several questions before we begin here:

(1) What should we use as boundary conditions for the side where $r \to \infty$?
(2) How do we cope with a need for widely varying step sizes, particularly with the modified model (and low-energy states of the unmodified model) where we need much more detail in a small region near the origin, and with the requirement[12] that H'(0)=K'(0)=J'(0)=0?
(3) How do we make sure of reasonable numerical accuracy?
(4) How do we get accuracy *both* in the function values *and* in their derivatives, so that we can get a reasonable approximation to the Hessian for use in analysis of the stability of the solutions?

For the boundary condition on H, a comparison of Prasad and Sommerfeld[25] with Hasenfratz and 'tHooft[3] shows that they are both calling for:

$$\frac{H(r)}{r} \to F \quad \text{as } r \to \infty, \tag{21}$$

where we may take H(∞) and F both as positive without loss of generality. A careful analysis of Julia and Zee[24] shows that we may use:

$$K(r) \to 0 \quad \text{as } r \to \infty \tag{22}$$
$$J(r) \to b \quad \text{as } r \to \infty \tag{23}$$

where b is any constant we may choose, representing the charge of the solution. For monopole solutions, we may simply pick b=0. (Julia and Zee also allow for a term "Mr," where M represents the voltage $A_0$ of the vacuum field at r=∞; we may treat it as zero, without loss of generality.)

To address the second question, we may build a numerical procedure or subroutine which inputs an entire array of user-specified stepsizes deltar(1:N), where N+1 is the number of grid points that we want to represent explicitly. For algebraic convenience, we may denote deltar(k) as $\Delta r_k$. Thus on each iteration, the numerical routine could update its estimates of $H(r_k)$, $J(r_k)$, $K(r_k)$, $H'(r_k)$, $J'(r_k)$ and $K'(r_k)$ for all k from 0 to N, where :

$$r_0 = 0 \tag{24}$$
$$r_k = r_{k-1} + \Delta r_k \quad \text{for k = 1 to N} \tag{25}$$

***However,*** for the sake of numerical accuracy near zero and near infinity, it is better to use a slightly different representation of the functions in the computer, still using the same ansatz. For all choices of k below some choice $k_0$ (perhaps N/2 or perhaps based on a subroutine input $r_0$), we may borrow an idea from Affouf[21], by actually updating the functions h(r), j(r) and k(r) defined by:

$$H(r) = rh(r) \tag{26}$$
$$J(r) = rj(r) \tag{27}$$
$$K(r) = 1 + rk(r) \tag{28}$$

More precisely, we may update both these functions and *their* derivatives with respect to r, on each iteration. It is easy enough to substitute equations 26 through 28 (and their r derivatives) into the energy term in equation 18, to see how to do this.

***Likewise,*** a modified representation allows greater accuracy for the upper half of the grid. The boundary conditions 21-23 suggest that we should represent the functions H, J and K as h(r), J(r) and K(r) for k≥$k_0$, and represent them as functions of (1/r) instead of r. This does not mean altering equations 24 or 25, or the locations of the grid point; however, it would be more convenient to update the derivatives of h, J and K with respect to 1/r, for k>$k_0$. More important, it affects how we will address question four.

For the third question, we may again follow Affouf[21].

Affouf notes that Julia and Zee used a simple explicit forwards integration method from r=0 to r=∞. Methods of that type are well-known to have problems with numerical instability. As an alternative, Affouf used a kind of Newton's method (using a sparse matrix solver) to "minimize" this function of the grid estimates.

A few details require some explanation. First, of course, Affouf used a different mapping from grid estimates to an estimate of $e^2E/4\pi$; he did not need to consider variable step sizes or the accuracy of derivatives as such. Second, Affouf proved the stability of his numerical method, but this is not the same as proving stability of the soliton itself. Third, Affouf did comment that prior numerical experiments with the Georgi-Glashow system were less rigorous and reliable numerically.

After finding solutions in this way, it is natural to assume that one has performed a minimization, and that the solutions themselves must be stable. However, it is well-known that solutions of the Lagrange-Euler equation may be minima, maxima, or saddle points. A key test is to look more closely at the (large, sparse, block-diagonal) Hessian matrix from the final stage of iteration, and simply run it through a sparse matrix solver for the lowest eigenvalue and corresponding eigenvector. If that eigenvalue is negative, one may even test the corresponding eigenvector as a perturbation to see that it does in fact reduce the energy of the system.

Given a candidate solution, it is possible to test for stability more quickly by simply using dynamic programming along the line from r=0 to r=∞. However, because we want to find a reliable solution to start from in any case, it should be easy enough to just run the Hessian through an eigenanalysis; this costs more computer time than dynamic programming, but may reduce the cost in terms of human labor.

Finally, for the sake of stability analysis, it is especially important that we seek accuracy both in the function values *and in* their derivatives with respect to r. In order to achieve this, we may use a *cubic spline interpolation* to interpolate the functions h, j and k as functions of r between grid points $r_{k-1}$ and $r_k$, for $k \leq k_0$. Likewise, we may use a cubic spline interpolation to interpolate the functions H, J and K as functions of 1/r, for values of r between grid points $r_k$ and $r_{k+1}$ for $k \geq k_0$. More precisely, before we do any computational work, we simply substitute these interpolations into equation 18, and work out what the integral equals between grid points $r_k$ and $r_{k+1}$. For the final grid interval, from $r_N$ to ∞, we may extrapolate h(r)-h(∞), J(r)-J(∞) and K(r) as an exponential decay; note, for example, we only need to know $K(r_n)$ and $K'(r_n)$ in order to fit a and b in the formula a*exp(-br). We use the Newton's method to minimize the sum of the resulting algebraic expressions, over all N+1 grid intervals.

For the case where the correction term in equation 19 is added, and where our Newton's method package is truly capable of minimization, the method may be further refined as follows. The values of $r_k$ calculated from equations 24 and 25 can be used as *initial values* for the grid point location. Then on each iteration, $r_k$ itself can be included, along with the three function estimates and three derivative estimates, as part of what is being estimated and optimized. This should work, because the algebraic expression we are minimizing is the *exact* integral of $(e^2/4\pi)E$ for the stepwise-cubic function being assumed. When we minimize this with respect to the choices of $r_k$, we are simply searching for the function values of minimum energy, throughout the entire space of piecewise-cubic functions with no more than N pieces.

After this kind of initial analysis, a logical next step is to modify equation 18 to account for spherical harmonics, so as to span the entire space of smooth perturbations. We may work out $\delta^2 E$ algebraically for equation 5, for the general case, and then insert the usual ansatz *only for* $Q^a$ and $A_\mu^a$, while allowing for more general possibilities for the perturbations. After considering perturbations within the usual ansatz (as above), the next obvious class of perturbations to check would be those of the form:

$$\delta Q^a = r_a \delta Q(r) \qquad (29)$$

$$\delta A_i^a = \varepsilon_{ial} r_l \cos\theta\, w(r) \tag{30}$$

From the discussion above, these are probably the perturbations most likely to expose instabilities here. If the adiabatic potential hypothesis is true, stable solitons in the modified tP model are likely to "spiral in" from the asymptotic boundaries, and require the use of spherical harmonics for an efficient, convergent numerical approximation. As with complex calculations in electronics, it may even be necessary to use adaptive basis functions and neural networks, in order to maximize accuracy subject to a constraint on the use of computer time.

## **Appendix B: P transforms and well-posedness of quantum field theory**

How can classical-to-quantum mappings[9,10,11] assist us in finding out whether specific quantum field theories are well-posed dynamical systems?

To begin with, they can play a crucial role in defining what it means to *be* a well-posed dynamical system. According to the many-worlds version of quantum field theory[13], the state of the universe is defined by a wave function $\Psi$ which obeys:

$$\dot{\psi} = iH\Psi \tag{31}$$

But what does it mean for this to be a well-posed dynamical system?

For technical reasons, it is more convenient here to express quantum field theory in terms of the usual density matrix $\rho$, which leads[13] to the usual dynamical equation:

$$\dot{\rho} = i[H, \rho] \tag{32}$$

We may then ask: when is equation 32 a well-posed dynamical system.

Even with classical PDE, there is no universal definition of what it means to be well-posed. However, it often possible to prove that a set of PDE, started out at time 0 with field values in the Sobolev space $W^{k,p}$, will have unique solutions which remain in this space for all times t. Here, I would propose that we consider the Sobolev spaces $PW^{k,p}$, which are simply the spaces spanned by the density matrices $P(\{\pi,\varphi\})$ for field states $\{\pi,\varphi\}$ which are in $W^{k,p}$. We may attempt to answer the following question: for various Hamiltonian operators H, and initial values of $\rho$ in $PW^{k,p}$, when can we be sure that there will exist a unique solution for $\rho$ at later times, also within $PW^{k,p}$?

The P mapping only works for bosonic field theories, like those discussed in this paper. (To address the case of mixed Fermi-Bose quantum field theories, I propose that we represent such theories as a limiting case of bosonic field theories[12]; for example, as $\eta \to 0$ in section 4.2, the limiting case of the solitons may be particles with their charge concentrated at a point, and with half-integral spin[3,12].) It would be interesting enough at this stage in our knowledge to prove well-posedness of equation 32, for bosonic field theories in which the Hamiltonian H is just the P-mapped version of the classical Hamiltonian.

More precisely, we may quantize the tP model as follows. First, express the classical Lagrangian in the Lorentz gauge, as was done in the initial quantization of electrodynamics decades ago. Next, write out the classical Hamiltonian density $\mathcal{H}(\underline{\varphi},\underline{\pi},\nabla\underline{\varphi})$ for this system. Third, insert the usual field operators, *mapping* the multiplication between fields into the normal product of the corresponding field operators. The well-known results on P mappings then tell us that the resulting Hamiltonian operator H has the property that Tr(H P($\underline{\varphi},\underline{\pi}$)) equals the classical energy H of the original field state {$\underline{\varphi},\underline{\pi}$}. This equivalence may be of great use in characterizing properties of density matrices ρ=P($\underline{\varphi},\underline{\pi}$) by reference to properties of the corresponding classical fields. Of course, we may then augment the Hamiltonian by adding the $\eta(\partial_\mu Q^a)^2$ term to the Lagrangian, as described in section 4.2, and redoing this.

It should be noted that the evolution of the probability distributions for {$\underline{\varphi},\underline{\pi}$} in forwards time will *not* in general match equation 32. However, according to the backwards time theory of physics[13], this is not the correct way to deduce the probabilities for what we actually observe at the macroscopic level. Equilibrium states and even scattering states should in principle be deduced from the quantum Boltzmann distribution, exp(-kH), which is exactly the same for the quantized model and for the original classical model, *so long as* we limit consideration to states in $PW^{k,p}$ and in $W_{k,p}$, respectively. (For scattering states, we may use the old machinery of considering statistics within a periodic space of volume V, and looking for the limit as V→∞.) If the classical PDE are well-posed, and we develop methods to perform the spectral and scattering calculations, we will still have a well-posed model of physics, even if 31 and 32 should not be well-posed.